\documentclass[runningheads]{llncs}
\usepackage[T1]{fontenc}
\usepackage{graphicx}
\usepackage{hyperref}
\usepackage{color}

\usepackage{orcidlink}
\usepackage[capitalize]{cleveref}
\usepackage{xspace}
\usepackage{bbding}

\newcommand{\codename}{MATCH\xspace}
\newcommand{\tagline}{Multi-Agent Transparent and Controllable Human-centered\xspace}

\newif\ifanonymized
\anonymizedfalse  

\begin{document}
\title{\codename: Engineering Transparent and Controllable Conversational XAI Systems through Composable Building Blocks}
\titlerunning{Building Blocks for Transparent and Controllable Conversational XAI}
%
\ifanonymized
    \author{Anonymized}
\else
    \author{
        Sebe {Vanbrabant}$^($\Envelope$^)$\,\orcidlink{0009-0001-7996-6048} \and
        Gustavo {Rovelo Ruiz}\,\orcidlink{0000-0001-7580-8950} \and
        Davy {Vanacken}\,\orcidlink{0000-0001-8436-5119}
    }
\fi
\ifanonymized
    \authorrunning{Anonymized}
\else
    \authorrunning{S. Vanbrabant et al.}
\fi
%
\ifanonymized
    \institute{Anonymized}
\else
    \institute{Hasselt University - Flanders Make, Digital Future Lab, Diepenbeek, Belgium
        \email{\href{mailto:sebe.vanbrabant@uhasselt.be}{sebe.vanbrabant@uhasselt.be}}
        \email{\href{mailto:gustavo.roveloruiz@uhasselt.be}{gustavo.roveloruiz@uhasselt.be}}
        \email{\href{mailto:davy.vanacken@uhasselt.be}{davy.vanacken@uhasselt.be}}
    }
\fi
\maketitle              
\begin{abstract}
While the increased integration of AI technologies into interactive systems enables them to solve an increasing number of tasks, the black-box problem of AI models continues to spread throughout the interactive system as a whole. Explainable AI (XAI) techniques can make AI models more accessible by employing post-hoc methods or transitioning to inherently interpretable models. While this makes individual AI models clearer, the overarching system architecture remains opaque. This challenge not only pertains to standard XAI techniques but also to human examination and conversational XAI approaches that need access to model internals to interpret them correctly and completely. To this end, we propose conceptually representing such interactive systems as sequences of \emph{structural building blocks}. These include the AI models themselves, as well as control mechanisms grounded in literature. The structural building blocks can then be explained through complementary \emph{explanatory building blocks}, such as established XAI techniques like LIME and SHAP. The flow and APIs of the structural building blocks form an unambiguous overview of the underlying system, serving as a communication basis for both human and automated agents, thus aligning human and machine interpretability of the embedded AI models. In this paper, we present our flow-based approach and a selection of building blocks as \codename: a framework for engineering \tagline systems. This research contributes to the field of (conversational) XAI by facilitating the integration of interpretability into existing interactive systems.

\end{abstract}
\section{Introduction}
\label{sec:introduction}
The growing complexity of AI models, from interpretable decision trees to opaque Deep Neural Networks (DNNs) and Large Language Models (LLMs), has led to a decline in their transparency~\cite{arrieta2020explainable,luo2024understanding}. These models are often black boxes, producing results without explanations, justifications, or indications of uncertainties~\cite{von2021transparency}. The field of eXplainable AI (XAI) addresses these challenges by complementing AI predictions with explanations~\cite{gunning2019xai,longo2024explainable,xu2023xair}. Machine learning (ML) workflows can be made more transparent in two main ways: either by using white-box models that offer inherent interpretability, like decision trees, or by leveraging post-hoc explanations (e.g., LIME~\cite{ribeiro2016should} and SHAP~\cite{lundberg2017unified}) to try to explain the internal workings of black-box models, such as neural networks.

While established methods can explain the behavior of individual models, they often fall short when it comes to explaining the broader interactive systems in which those models operate. This is a significant limitation, as the quality of users' mental models of those systems directly influences how effectively they can interact with them~\cite{kulesza2012tell}. We see similar challenges with LLMs, which require careful prompting and the right (amount of) information to respond accurately and avoid hallucinations~\cite{huang2025survey}. To address the challenges of understanding the broader interactive systems that embed AI technologies, it is important to rethink how we approach the engineering of such interactive systems~\cite{dix2023engineering}.
                                                  
In the context of engineering interactive systems that incorporate LLMs, recent approaches like Tool-Augmented Language Models (TALMs)~\cite{parisi2022talm} and the Model Context Protocol (MCP)~\cite{mcp} allow LLMs to interface with external tools, supporting their integration into such systems. Furthermore, popular pipeline-based approaches like scikit-learn's pipelines and LangChain's chains provide structured ways to compose workflows of AI models and various utilities. However, these approaches remain agnostic to the user interface (UI) of the broader interactive system, which is often a conventional graphical user interface (GUI), resulting in only loosely connected conversational agents and GUIs, and thus interactive systems that fail to capitalize on the full potential of combining different UI paradigms~\cite{nielsen2023paradigm,ziegler2025challenges}.

In this paper, we aim to take an essential step towards engineering interactive systems with a deeper integration of conversational agents and GUIs, with the specific goal of aligning human and machine interpretations of complex systems that embed AI technologies. To this end, we introduce \codename, a framework for engineering \tagline systems. \codename makes AI workflows accessible and controllable for agents (including both users and automated agents, such as LLM-based agents) through \textbf{structural} and \textbf{explanatory building blocks}. For instance, an AI model (a structural building block) can be explained using techniques like LIME~\cite{ribeiro2016should}, SHAP~\cite{lundberg2017unified}, and WhatIf~\cite{wexler2019if} (explanatory building blocks). Additionally, the AI pipeline can be further controlled through structural building blocks that override decisions~\cite{kieseberg2023controllable} or provide per-instance feedback~\cite{kulesza2015principles}. We identify a selection of explanatory building blocks that address common user questions in interactive systems~\cite{lim2009assessing,lim2010toolkit,lim2019these} and structural building blocks that support direct control of AI behavior. We incorporate these building blocks in a common pipeline-based approach, which we extend into the interactive layer by exposing individual blocks as a common knowledge base. This enables both human and automated agents to audit system behavior, which is critical for engineering responsible AI systems that are both safe and trustworthy~\cite{stumpf2025phawm}. In doing so, we address the common knowledge base highlighted by Ziegler~\cite{ziegler2025challenges}.

Numerous approaches to conversational XAI exist~\cite{garofalo2025conversational,he2025conversational,kuzba2020what,malandri2023convxai,nguyen2023from,slack2023explaining,vanbrabant2025echo}. This work centers on the engineering challenges of seamlessly integrating such approaches into existing AI workflows, combining them with GUI elements and visual explanations in a non-invasive way. To ensure the generalizability of our work, we avoid imposing strict assumptions on the structure of AI-related code, which may run in diverse environments, from experimental notebooks to systems in production. Furthermore, unlike existing approaches that typically offer explanations and insights about model behavior, our block-based framework also enables control over model behavior and execution flow. Importantly, our goal is not to replace visual XAI methods with a purely conversational approach, but to integrate both paradigms side by side. This work elaborates on the preliminary research by Vanbrabant et al.~\cite{vanbrabant2025composable}, with the following engineering contributions:
\begin{itemize}
    \item \textbf{C1}: A conceptual 5-layer architecture that defines how structural and explanatory building blocks can be organized to engineer transparent and controllable interactive AI systems. The architecture provides an auditable interface accessible to multiple agents, including human and automated agents.
    \item \textbf{C2}: An implementation of this architecture in the form of \codename, a modular framework that exposes developer-written AI code as composable building blocks and connects them to conversational and GUI-based interfaces.
    \item \textbf{C3}: A catalog of structural and explanatory building blocks, grounded in prior work, that systematically capture common forms of control and explanation in interactive AI systems.
\end{itemize}

This paper first gives an overview of relevant concepts and approaches from related work in \cref{sec:related-work}. In \cref{sec:engineering}, we link these together in \codename, and we provide a catalog of structural and explanatory building blocks in \cref{sec:bb}. Finally, we reflect on our work and future directions in \cref{sec:conclusions}.

\section{Related Work}
\label{sec:related-work}
This section explores approaches for structuring and interacting with AI systems. We examine pipeline-based and neuro-symbolic AI methods, as well as flow-based interfaces that explicate the structure and training of AI models as an inspiration for our work.

\subsection{Pipelines-Based and Neuro-Symbolic AI}
In a typical AI workflow, two major stages can be identified: a data-oriented stage, involving data preparation, and a model-oriented stage, involving model (re)training and deployment~\cite{amershi2019software}. For supervised ML, the workflow results in a model that can predict new outputs from new inputs by leveraging its internal learning process. We can thus conceptually view a trained model as a pipeline that transforms inputs into outputs through a model. Pipelines can be chained to make system behavior more advanced and fit for a task. This pipeline-based view can be expanded to include multiple utilities. For instance, scikit-learn\footnote{\scriptsize\url{https://scikit-learn.org/stable/modules/compose.html}} supports pipelines to compose transformers for data preprocessing and predictors for ML models. Similarly, the LangChain framework\footnote{\scriptsize\url{https://github.com/langchain-ai/langchain}} is built around chaining LLMs with various utilities, such as input prompts and structured output formatters.

Pipeline-based approaches facilitate integrating both symbolic and neural methods into a larger system. \textit{Symbolic AI}, such as decision rules, excels at structured reasoning and provides inherent explainability and interpretability~\cite{wang2025towards}. However, symbolic approaches are less trainable and more error-prone in unfamiliar situations. In contrast, \textit{connectionist} techniques like neural networks excel at training by discovering and learning patterns from data, yet require large datasets for effective training and act as black boxes regarding explainability. \textit{Neuro-symbolic AI (NSAI)} integrates neural and symbolic approaches to combine the individual strengths of trainability and interpretability, circumventing their inherent weaknesses~\cite{wang2025towards}. Type 2 NSAI, as described by Kautz~\cite{kautz2022third}, considers connectionist models as neural module subroutines within a symbolic problem-solving system; the aforementioned TALMs are a recent example. TALMs query tools rather than generating answers directly, which is helpful for mathematical operations or to interface with external APIs. Furthermore, systems like ViperGPT~\cite{suris2023vipergpt} and Chameleon~\cite{lu2023chameleon} use LLMs as neural subroutines within a symbolic tool usage framework for visual questioning answering. 

\subsection{Conversational and Flow-Based Interfaces for Explainable AI}
LLMs offer interesting opportunities for XAI, as demonstrated in x-[plAIn]~\cite{mavrepis2024xai} and SHAPstories~\cite{martens2025tell}, which generate audience-specific summaries of XAI methods. These summaries are tailored to the users’ knowledge and interests, improving accessibility and decision-making. Such approaches, however, do not offer capabilities beyond those of the integrated XAI methods. ECHO~\cite{vanbrabant2025echo} takes the conversational approach to XAI one step further, with a TALM using generated tools to explicate system-specific behavior, complemented with predefined tools that address various explanation types and XAI methods. Nevertheless, these approaches are purely textual, and would benefit from an extension with intelligible visual approaches. This works both ways: visualizations like those in TimberTrek~\cite{wang2022timbertrek} and AI-Spectra~\cite{eerlings2024ai} could be complemented with conversational interfaces to help users understand and select the right models for their needs.

A number of tools rely on flow-based, graph-like visualizations. For instance, DeepGraph~\cite{hu2018deepgraph} visualizes models during
development by constructing a data flow graph of the architecture from the DNN source code and automatically synchronizing it with a graph representation. DeepFlow~\cite{calo2025deepflow}, on the other hand, uses flow-based visual programming to realize a no-code approach to building neural networks by viewing models as sequences of learnable functions. Such approaches help visualize complex, internal architectures, but primarily focus on the development process rather than the model's role within the encapsulating system. In contrast, \codename extends the scope of pipeline- and flow-based approaches for model construction and training by integrating explainability, controllability, and interaction within a unified framework. In doing so, \codename addresses not only the technical aspect of composing AI workflows, but also the engineering challenge of making these workflows intelligible and controllable.

\section{\codename: A Framework for Engineering \tagline Systems}
\label{sec:engineering}
Current explainability methods typically probe AI models by analyzing parts of the \textit{input $\rightarrow$ model $\rightarrow$ output} pipeline. We expand this view by introducing building blocks that enable AI models to be systematically visualized and controlled within the larger, embedding AI system. Inspired by type 2 NSAI, we represent the AI workflow through \textit{structural building blocks} (e.g., a neural network) that offer both human and automated agents a common knowledge base about the architecture. Each structural building block can then be further examined through \textit{explanatory building blocks} (e.g., LIME/SHAP).

To illustrate our approach, we use the publicly available loan prediction dataset\footnote{\scriptsize\url{https://www.kaggle.com/datasets/altruistdelhite04/loan-prediction-problem-dataset}}, a common use case in critical decision-making~\cite{green2019principles,he2025conversational}. This section outlines our conceptual 5-layer architecture for engineering these building block-based systems, as well as the corresponding \codename framework.

\subsection{Auditable 5-Layer Architecture for Engineering Conversational XAI Systems Using Composable Building Blocks}
\begin{figure}[t]
    \centering
    \includegraphics[width=1\linewidth]{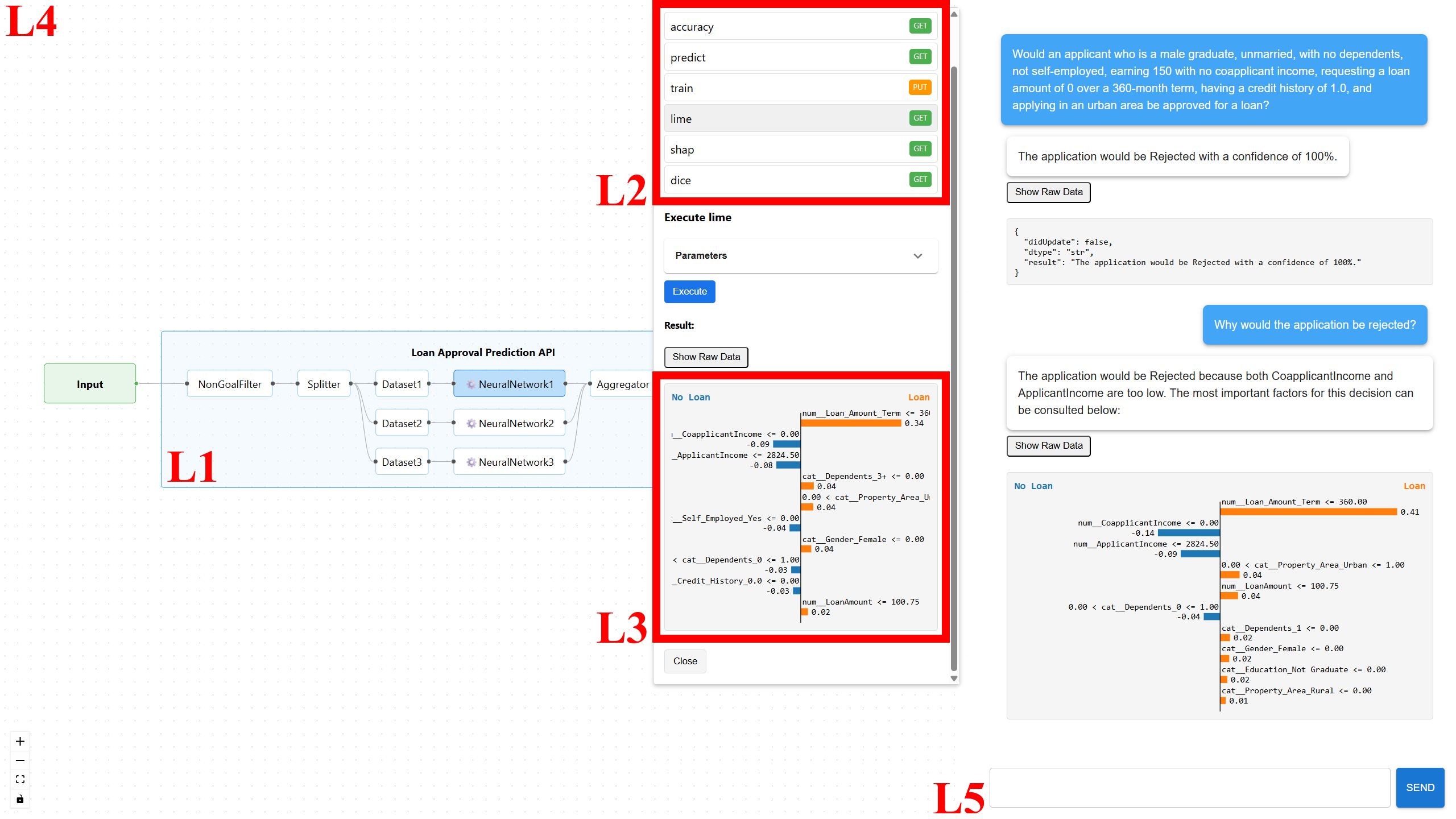}
    \caption{Prototype of our approach, using the loan approval prediction ensemble. The UI (layer~4) shows each structural building block (layer~1) influencing predictions, exposing system behavior to both human and automated agents (layer~5) through explanatory building blocks (layer~3) and an API (layer~2), respectively. The image shows a LIME explanation for both the NeuralNetwork1 network and the total chain. Both the UI and LLM have access to the same API and are able to use the same visual components.}
    \label{fig:usecase}
\end{figure}

\cref{fig:architecture} shows our proposed 5-layer architecture, using an example ensemble trained on the loan approval prediction dataset. The vertical pipeline is loosely based on the XAI system by Mohseni et al.~\cite{mohseni2021multidisciplinary}, modified to support structural and explanatory building blocks. We define the following layers, which are also depicted in \cref{fig:usecase}:

\begin{figure}[t]
    \centering
    \includegraphics[width=1\linewidth]{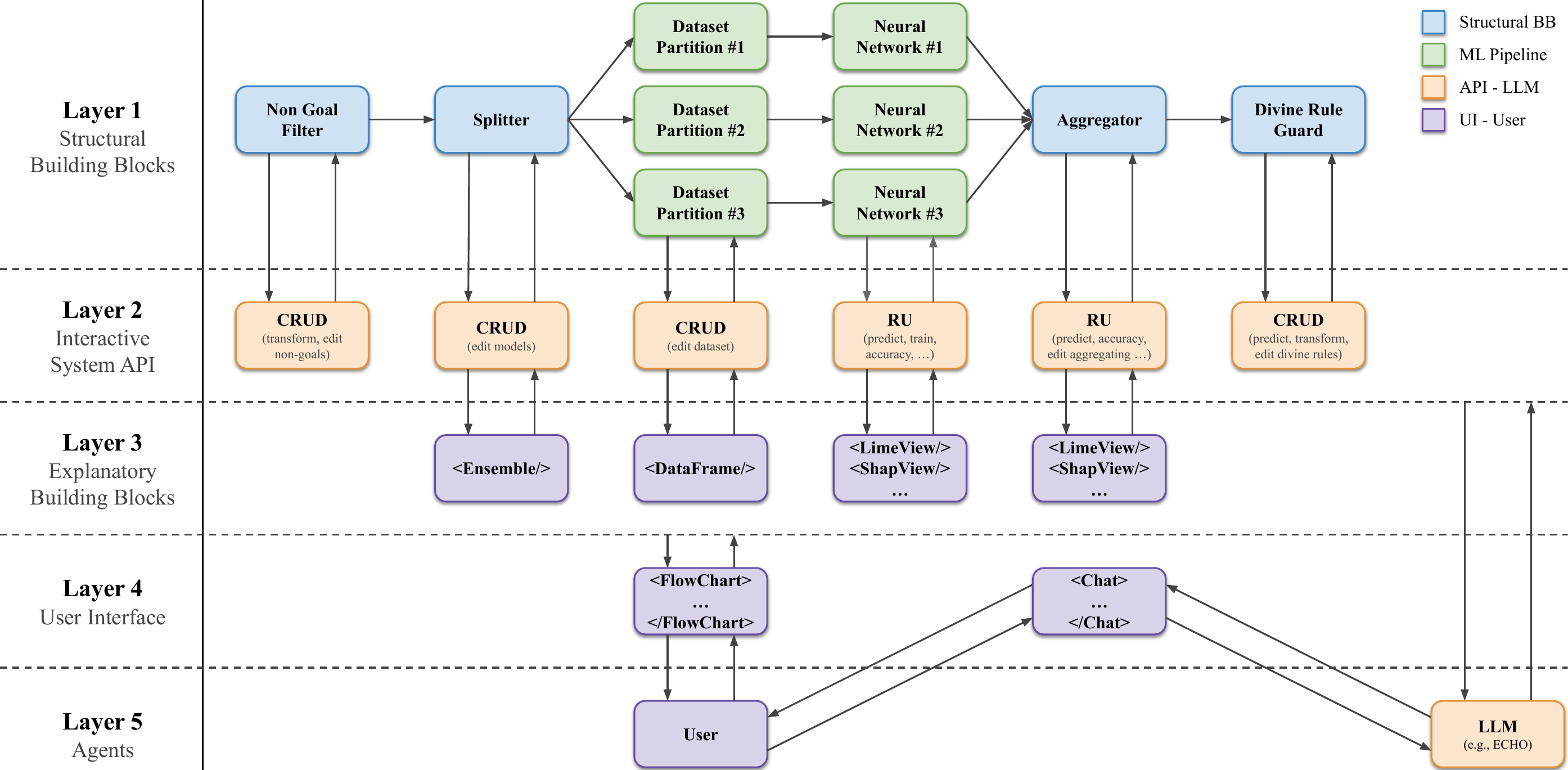}
    \caption{An example loan approval prediction ensemble to illustrate our proposed architecture. Layer~1 shows the structural building blocks, consisting of our building blocks (in blue) integrated into the ML pipeline (in green). Layer~2 converts the structural building blocks into a callable API, usable by the explanatory building blocks in layer~3, assembled into the UI of layer~4. The LLM of layer~5 can interact with the underlying system via the API of layer~2, so that all agents operate on a common knowledge base. Agents can communicate among themselves via a chat interface.}
    \label{fig:architecture}
\end{figure}

\begin{description}
    \item[Layer~1: Structural Building Blocks.]{These blocks convey the (conceptual) architecture of the interactive system. Each block is mapped to a part of the system's source code, as specified by the developer via functions and decorators (see \cref{sec:building-block-language}). Since this representation is conceptual, developers can choose what (parts of) the underlying system to expose and how to structure/communicate the pipeline.}

    \item[Layer~2: Interactive System API.]{Where the conceptual building blocks of layer~1 structure how developer-defined functions relate to structural components, layer~2 transforms each block's functions into an API that conveys their scope and purpose to the front-end UI and automated agents. To enable interactions, these methods are converted into a REST API. \cref{sec:api-generation} details how the API generation process is put into practice.}

    \item[Layer~3: Explanatory Building Blocks.]{These interact with the structural building blocks of layer~1 through the API of layer~2. For instance, a structural building block of an AI model can have its behavior explained through LIME. Similarly, the model's training dataset can be explored through a DataFrame\footnote{\scriptsize\url{https://pandas.pydata.org/docs/reference/api/pandas.DataFrame.html}} visualization.}

    \item[Layer~4: User Interface.]{Structural and explanatory building blocks are combined into a UI where they can be explored, interrogated, and controlled. Currently, explanatory building blocks are visualized and assembled into a coherent UI or as part of chat messages; future research directions include LLM-generated layouts that dynamically adapt to user preferences~\cite{brie2023evaluating}.}

    \item[Layer~5: Agents.]{The final layer includes the agents that interact with the building blocks. These can be users interacting with the UI and its explanatory building blocks, or an automated LLM agent interacting with the shared API of layer~2. This API can then be integrated as a set of tools for a TALM, such as ECHO~\cite{vanbrabant2025echo}. Both agents have access to the same knowledge base, allowing users, for instance, to query the LLM about system behavior.}
\end{description}

\subsection{Implementing \codename: Building Block Language and Multi-Agent API Integration}
We designed \codename as a lightweight integration framework that exposes existing AI workflows as composable and inspectable building blocks, as detailed in our 5-layer architecture. The central idea is to preserve the developers’ existing codebase while making its internal logic auditable and manipulable through a unified interface. For instance, we assume the loan prediction AI system already exists; instead of refactoring the existing code to fit our framework, \codename simply provides a structured way for developers to map existing functionality onto conceptual building blocks (e.g., a \textit{neural network} block with a \textit{predict} method). These structural building blocks clarify the methods' purpose and scope and allow \codename to expose the system's codebase through a generated API.

This addresses a key limitation of the ECHO framework~\cite{vanbrabant2025echo}, which places strict requirements on the definition of the AI models with which it can interact. In contrast, \codename only requires developers to define which methods to expose, link them to conceptual building blocks to define their scope, and chain building blocks together to represent the overall system.

\subsubsection{Building Block Language.}
\label{sec:building-block-language}
In \codename, the methods that the developer wants to expose are first linked to \texttt{BuildingBlock}s by passing them to the blocks as method lists of \texttt{Callable}s (the syntax is described in EBNF in \cref{fig:ebnf}). Next, these building blocks are linked together in a chain that explicates the system as a pipeline. As described in \cref{sec:related-work}, constructing AI workflows as pipelines is a widely adopted paradigm.

\texttt{BuildingBlock}s can be chained sequentially (denoting input-output flow) or in parallel (denoting multi-model ensemble-like approaches). In our Python implementation of \codename, sequential chaining is facilitated by overloading the \texttt{\_\_or\_\_} and \texttt{\_\_ror\_\_} operators (pipe operator `\texttt{|}') on the \texttt{BuildingBlock} class, which chains two blocks together into an instance of the \texttt{Chain} class. To chain blocks in parallel, individual \texttt{Runnable}s are passed as \texttt{ParallelBlock} lists. A \texttt{Runnable} can either be a \texttt{BuildingBlock}, \texttt{Chain}, or \texttt{ParallelBlock}. This approach is inspired by the LangChain framework and allows developers to express AI workflows in a concise and uniform way.
\begin{figure}[ht]
    \begin{verbatim}
Runnable      ::= BuildingBlock | Chain | ParallelBlock

BuildingBlock ::= "BuildingBlock" "(" name, "," MethodList ")"

MethodList    ::= "[" Callable* "]"

Chain         ::= Runnable "|" Runnable

ParallelBlock ::= "ParallelBlock" "(" Runnable ("," Runnable)+ ")"
    \end{verbatim}
    \caption{Grammar in EBNF for defining workflows consisting of building blocks in \codename. The AI codebase is linked to individual \texttt{BuildingBlock}s as lists of \texttt{Callable}s (i.e., methods). Each type of building block is a \texttt{Runnable} that can be chained sequentially via pipeline operators, or in parallel via \texttt{ParallelBlock}s.}
    \label{fig:ebnf}
\end{figure}

\subsubsection{Exposing Building Blocks to Agents.}
\label{sec:api-generation}
\begin{figure}[ht]
    \centering
    \includegraphics[width=1\linewidth]{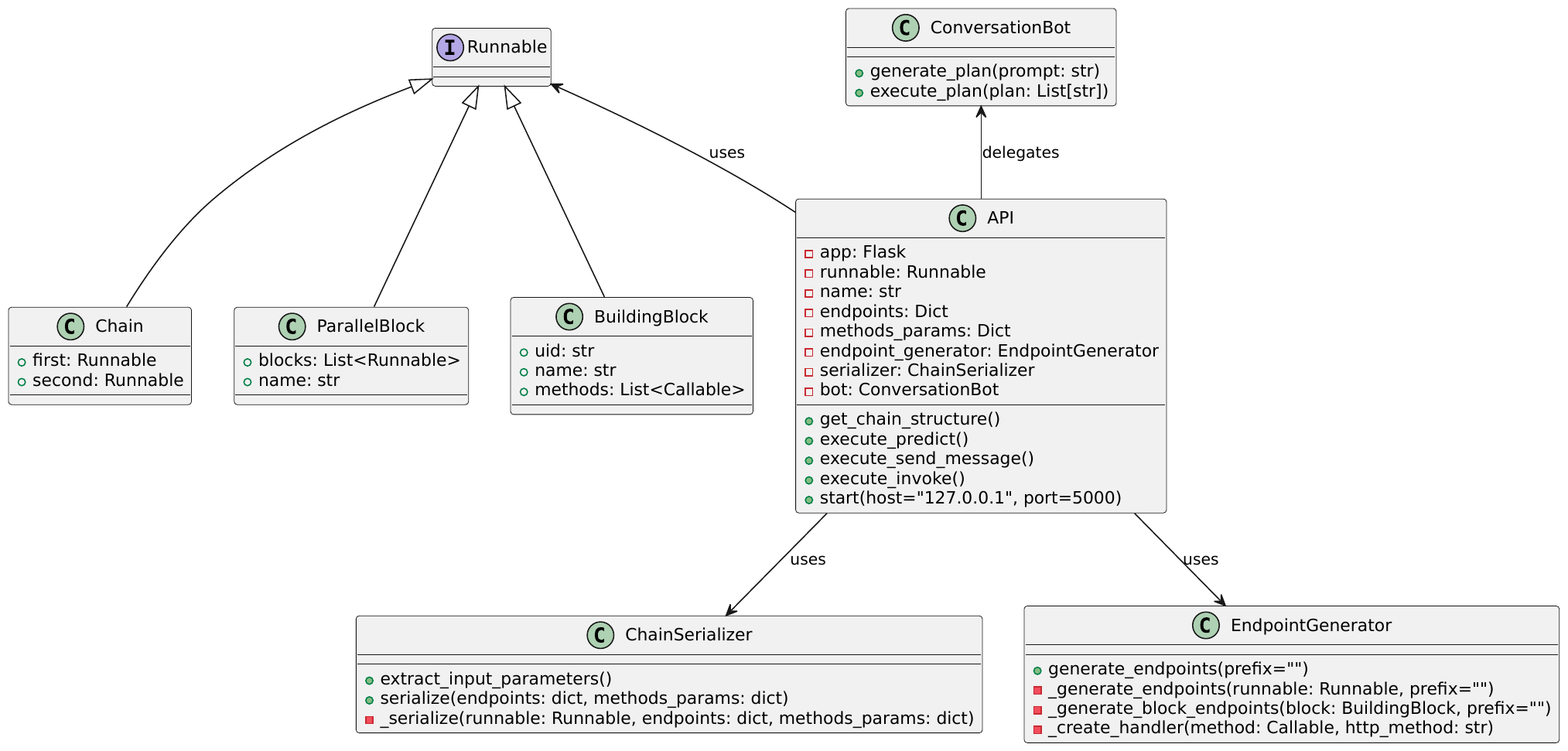}
    \caption{Diagram of the dynamically generated API. The \texttt{API} class recursively traverses a \texttt{Runnable}, which can be a \texttt{BuildingBlock}, \texttt{Chain}, or \texttt{ParallelBlock}, to expose their methods as REST endpoints via the \texttt{EndpointGenerator} utility class. Methods annotated with CRUD or prediction decorators are automatically mapped to HTTP methods (POST, GET, PUT, DELETE). The structure of the entire \texttt{Runnable} is serialized via the \texttt{ChainSerializer} class. This workflow enables users to interact with the underlying AI workflow via a UI using the API, while the \texttt{ConversationBot} (in our case, ECHO~\cite{vanbrabant2025echo}) enables conversational access via generated tools.}
    \label{fig:plantuml}
\end{figure}

To integrate the building blocks in a front-end UI and enable conversational interactions, \codename auto-generates a REST API of the building blocks, exposed via Flask\footnote{\scriptsize\url{https://github.com/pallets/flask}}. \codename also leverages ECHO's tool generation process for conversational XAI for the back end~\cite{vanbrabant2025echo}, summarized in \cref{fig:plantuml}.

The \texttt{API} class recursively traverses a \texttt{Runnable}, generating endpoints for each \texttt{BuildingBlock}'s methods that are marked with CRUD (i.e., Create, Read, Update, Delete) decorators. Such decorators are used in \codename to attach additional metadata to methods. One key use of the associated CRUD metadata is to transform methods into the correct HTTP methods for the front-end-facing API and enable automatic propagation of updates through the pipeline. For example, if a dataset is modified via a method marked as \texttt{@bb\_update}, any downstream building blocks (e.g., AI models whose training process depends on the dataset) will have their methods marked as \texttt{@bb\_update} called automatically. Thus, updating a dataset might trigger retraining AI models further downstream automatically. Each endpoint enforces type-safe parameter parsing, executes the associated method, and serializes outputs to JSON, including metadata such as the data type and whether an update occurred. Based on this data type, the frontend and the automated agents know how to interpret a method's output.

The API also exposes special endpoints to fetch the overall chain structure, make predictions, and initiate a conversation with ECHO, enabling both human and automated agents to query, invoke, or send messages to the chain while operating on a common knowledge base. Internally, \codename identifies the chain’s predictive method, ensuring that calls to \texttt{predict} correctly route inputs through the sequential or parallel blocks. These methods are uniquely identified by marking them as \texttt{@bb\_predict}, which allows the framework to treat a chain as a functional pipeline by identifying which methods to call with which input parameters. So when a chain's \texttt{predict} method is called, inputs are passed sequentially through all \texttt{@bb\_predict} methods, transforming them into a final prediction. \codename propagates outputs of building blocks to inputs of subsequent building blocks, leaving the specific implementation and data handling of methods open to the developer. Additionally, input transformations can be triggered separately from predictions by methods marked with the \texttt{@bb\_transform} decorator. This is useful, for instance, for aggregating data after retrieving it from a \texttt{ParallelBlock}, or adjusting user input to conform to intended behavior (see \cref{sec:bb-control,sec:bb-flow}).

This design links the chain of building blocks directly to a dynamic front end, supporting both automated reasoning by automated agents such as LLMs and interactive use by human users without requiring modifications to the underlying developer-written code. While \codename offers a novel, structured approach to interacting with AI models and pipelines by converting methods into \textit{generated tools}, it could be further enhanced with \textit{predefined tools}, such as the reasoning tools in Chameleon~\cite{lu2023chameleon} or general-purpose intelligibility tools in ECHO~\cite{vanbrabant2025echo}.

\section{Structural and Explanatory Building Blocks}
\label{sec:bb}
This section presents an overview of possible structural and explanatory building blocks that can be integrated into the \codename workflow. Explanatory blocks help human and automated agents interpret AI behavior, while structural blocks convey the interactive system's structure and capabilities. Structural building blocks convey a \textit{conceptual} overview and interface of the underlying AI system, rather than imposing a strict framework. These blocks primarily use the \texttt{@bb\_transform} decorator so they can modify input to a \texttt{Runnable}; by implementing CRUD operations as well, agents can further customize how the input gets modified. We illustrate our building blocks on tabular data, in accordance with the loan approval use case, although \codename's modular workflow can be extended to other data types as well.

\subsection{Structural Building Blocks for Execution Flow}
\label{sec:bb-flow}
We first define building blocks for visualizing and controlling (conditional) execution flows. This is particularly useful when multiple AI models are used together, such as in the context of ensemble learning~\cite{opitz1999popular}, model multiplicity~\cite{eerlings2024ai,wang2022timbertrek}, or cascaded approaches~\cite{thys2025improving}, which are common in high-stakes interactive systems. Aside from the AI model and its dataset, we define a \textbf{Splitter} and \textbf{Aggregator}:

\begin{description}

    \item[Splitter]{routes input data to multiple models, enabling pipelines where multiple models are used in parallel, for example, each handling different sub-tasks or data segments, or providing redundant, alternative perspectives. Users can define routing rules through a visual interface, specifying how data is directed to particular models, and optionally augmented before model training.}

    \item[Aggregator]{aggregates the outputs from multiple models into a final decision. Users can configure aggregation strategies, such as majority voting, via a visual interface. The aggregator dashboard can display intermediate outputs and confidence levels, enhancing transparency in ensemble systems. In the context of model multiplicity, the AI-Spectra dashboard can visualize the aggregator's decision process, while Chernoff bots --- robot-like adaptations of Chernoff faces~\cite{chernoff1973faces} showing multivariate model configurations --- can represent each individual model~\cite{eerlings2024ai}.}

\end{description}

\subsection{Structural Building Blocks for Control}
\label{sec:bb-control}
In addition to explicating the flow between individual building blocks, we also want to make this flow, as well as the individual AI models, more controllable. For this, we primarily draw inspiration from the work on controllable AI by Kieseberg et al.~\cite{kieseberg2023controllable} and map their five methods for managing control loss onto the structural building blocks in our interactive pipeline as follows:

\begin{description}

    \item[DivineRuleGuard]{enforces ethical compliance by overriding harmful or unethical model outputs for non-autonomous AI systems. Users can interact with this block through a visual rule editor, which lets them define and activate conditions under which outputs must be overridden, while automated agents can generate such rules from natural language descriptions. In the loan approval context, rules could override a rejection when an applicant satisfies key financial thresholds, regardless of the model's prediction.}

    \item[NonGoalFilter]{acts as a pre-processor, rejecting inputs that do not align with intended behaviors or intentions. Similarly to the \textbf{DivineRuleGuard}, users can interact with this block through a visual rule editor. In the loan approval context, this could be used to reject inputs where essential fields are missing or fall outside plausible ranges (e.g., negative income), preventing invalid predictions from being generated.}

    \item[ShutdownTrigger]{functions as an emergency stop to disable autonomous AI systems at any point. While crucial for autonomous systems, it is less relevant for interactive human-in-the-loop systems where a human makes the final decision, such as our loan approval use case.}

    \item[IntentionalBiasInjector]{strategically influences decision-making by embedding corrective signals to guide the model toward preferred outcomes. Users can re-label undesirable predictions and submit these corrections to influence future outputs. For neural networks, this can be a lightweight adaptation approach inspired by LoRA and FiLM~\cite{hu2021lora,perez2018film}, where only selected layers of the neural network are updated or modulated. This allows efficient personalization and customization of AI behavior without retraining the full model.}

    \item[LogicBomb]{operates as a self-monitoring fail-safe, resetting or shutting down the AI if it ever attempts to produce an outcome that breaches ethical or operational boundaries. \textbf{LogicBomb} combines elements of \textbf{\mbox{NonGoalFilter}} and \textbf{ShutdownTrigger}, automatically resetting or shutting down the system based on triggered rules. This component can also monitor explanations, for instance, to guard for predictions when an explanation reveals it was mainly based on protected attributes like gender.}

\end{description}

These control-oriented building blocks also serve a broader purpose: they represent mechanisms to calibrate the equilibrium between automated decisions and human oversight. For instance, the \textbf{DivineRuleGuard} allows human stakeholders to override automated decisions in ethically sensitive cases, ensuring that automation supports rather than replaces human judgment. Similarly, the \textbf{\mbox{ShutdownTrigger}} ensures humans remain in control of the system by providing an emergency stop to halt automated decisions.

\subsection{Explanatory Building Blocks for AI Model Interpretability}
\label{sec:bb-explanatory}
We organize our explanatory building blocks around the commonly used explanation types of \textit{Why}, \textit{Why-not}, \textit{What-if}, \textit{How-to}, \textit{When}, and \textit{What-else}~\cite{lim2019these,mohseni2021multidisciplinary,vanbrabant2025echo}, which can be invoked straightforwardly in a conversational setting. For each type of explanation, we illustrate how established XAI methods can help in the loan approval context, and we give examples of how using explanatory and structural building blocks together can lead to more powerful systems.

\begin{description}

    \item[Causal Explanations (Why and Why-not)]{describe the reasoning behind how specific inputs lead to certain outputs. For black-box models, such explanations can be provided through explanatory LIME~\cite{ribeiro2016should} and SHAP~\cite{lundberg2017unified} blocks, which use feature importance to address \textit{Why} and \textit{Why-not} questions~\cite{lim2019these}. In our loan approval example, shown in \cref{fig:usecase}, consider a user wondering why their application was denied: a \textit{Why-not}-explanation would highlight the features that most contributed to their prediction. The explanation of the system pipeline can then be cross-referenced by inspecting the individual structural building blocks (e.g., AI models or rules/filters) as they might further impact the decision or the explanations of each individual model in a multi-model setup.}

    \item[Transfactual Explanations (What-if)]{allow users to simulate how the system responds to user-defined input values~\cite{lim2019these}. For \textit{What-if} questions, explanatory building blocks can display the input parameters of the \textit{predict} method and a corresponding output value, following the approaches of both He et al.~\cite{he2025conversational} and Vanbrabant et al.~\cite{vanbrabant2025echo}. In the loan approval example, \textit{What-if}-explanations enable users to explore hypothetical scenarios that might lead to a more desirable outcome. They can, for instance, test alternative input values based on the features highlighted by \textit{Why}-explanations. This can, in turn, be cross-referenced with structural building blocks that might impose further constraints on decision-making. Here, an automated agent can reason over all values and building blocks to provide holistic insights via \codename's API.}

    \item[Counterfactual Explanations (How-to)]{specify what changes to the inputs would yield a different output. While various methods exist~\cite{wachter2018counterfactual}, we adopt the DiCE framework~\cite{mothilal2020explaining}, which generates diverse and plausible counterfactuals using a model-agnostic approach that supports user-defined constraints and multiple target outcomes. We visualize counterfactuals by showing original inputs and modified inputs side-by-side, highlighting the changes that altered the prediction. \textit{How-to}-explanations automatically find out how users could change their loan approval status. Whereas established XAI techniques can generate counterfactuals for individual models, the API exposed by \codename enables users and automated agents to further reason about the logic in other structural building blocks to obtain system-level counterfactuals that reflect the behavior of the entire interactive pipeline.}

    \item[Generalized Prediction Conditions (When)]{reveal the typical conditions under which a prediction is made. Lim et al.~\cite{lim2019these} use \textit{prototypes} to represent inputs that reliably lead to a given outcome, and \textit{criticisms} to highlight atypical cases, helping users understand the generalizability and limits of the model’s behavior. We obtain prototypes and criticisms via the \mbox{MMD-critic} method of Kim et al.~\cite{kim2016examples}, and display them in a table. When working with interpretable models like decision trees, their internal structure can be shown instead~\cite{vanbrabant2025echo}. In the loan approval context, \textit{When}-explanations help users identify and understand the typical profiles that lead to approval or rejection, such as loans being approved above a certain income threshold.}

    \item[Explanations by Example (What-else)]{look for samples in a model’s training dataset that are similar to a given input in the model representation space that produce the same or similar outputs~\cite{mohseni2021multidisciplinary}. We display \textit{Examples} in a ranked list, ordered by similarity, with key features and prediction scores shown per instance. For loan applications, \textit{What-else}-explanations show similar historical cases, revealing to users how comparable applicants were treated by the system. This allows them to understand whether their profile typically leads to approval or rejection, and identify the key differentiating factors among similar applications.}

\end{description}

Integrating explanatory building blocks into \codename offers two key advantages over standalone use. First, as discussed in \cref{sec:introduction}, established XAI methods often fall short when it comes to explaining the broader interactive systems in which AI models are embedded. With \codename, these explanatory building blocks can also be applied at the system level (which is itself also a predictive \texttt{Runnable}), as shown in \cref{fig:usecase}. Second, once AI models are integrated into interactive systems, both the models and their accompanying XAI methods often become inaccessible. In contrast, \codename's conceptual representation of complex systems keeps the embedded AI technologies accessible and interrogable to users and automated agents alike.

\section{Conclusions}
\label{sec:conclusions}
Existing XAI techniques often fail to scale once integrated into larger workflows and complex interactive systems. As a result, challenges regarding escalating complexity and lack of transparency persist, not only at the level of the end user but also from an engineering perspective. We addressed this with \codename, a framework for engineering \tagline systems. \codename enables developers to expose their codebase through structured pipelines by chaining simple building blocks (\textbf{C2}). These structural and explanatory building blocks are integrated through an interactive systems API, situated in a conceptual 5-layer architecture. Users and automated agents such as LLMs can interact with the interactive system through GUI components or the API, respectively (\textbf{C1}). By explicitly communicating the pipeline of the interactive system, our approach provides both users and automated agents with a concrete mental model of the underlying architecture and enables them to unambiguously reason about complex pipelines.

Our catalog of building blocks is grounded in the literature: structural building blocks convey the structure of the interactive system, while explanatory building blocks offer explanatory insights into the structural components (\textbf{C3}). Through standardized CRUD operations, both users and automated agents can interrogate and control individual components via the GUI or via tool-calling in the context of TALMs. This combination of a transparent structure, explanatory views, and interactive control supports the engineering of multi-agent interactive systems that are both more transparent and more controllable compared to current pipeline-based approaches. Although we set our current focus to tabular data, \codename's modular approach can be extended to other data types as well.

\codename can benefit multiple stakeholders in distinct ways. Firstly, developers gain a minimally invasive mechanism to expose system functionality as composable building blocks. Secondly, designers can reason about the underlying system architecture through visualized workflows, facilitating the integration of human-centered design practices. Thirdly, end users can interact with explanations and control mechanisms, enabling them to interrogate and influence system behavior directly. By supporting multiple perspectives, \codename contributes to a more comprehensive engineering approach to embedding AI technologies in interactive systems. Furthermore, the modular nature of \codename makes it straightforward to add additional building blocks or conceptual layers that can further target diverse stakeholders.

\ifanonymized
\else
    \begin{credits}
    \subsubsection{\ackname}
    This work was funded by the Special Research Fund (BOF) of Hasselt University, BOF23OWB31.
    
    \subsubsection{\discintname}
    The authors have no competing interests to declare that are relevant to the content of this article.
    \end{credits}
\fi
%
%
%
\bibliographystyle{splncs04}
\bibliography{references}
\end{document}